\documentclass{ieeeaccess}
\usepackage{cite}
\usepackage{amsmath,amssymb,amsfonts}
\usepackage{algorithmic}
\usepackage{graphicx}
\usepackage{textcomp}
\def\BibTeX{{\rm B\kern-.05em{\sc i\kern-.025em b}\kern-.08em
    T\kern-.1667em\lower.7ex\hbox{E}\kern-.125emX}}

\usepackage{dcolumn}
\usepackage{bm}
\usepackage{braket}
\usepackage[normalem]{ulem}
\usepackage{comment}
\usepackage[export]{adjustbox}
\usepackage{booktabs, multirow}
\usepackage{fdsymbol}
\usepackage{url}
\usepackage{soul}
\sethlcolor{electriclime}

\usepackage{listings}
\lstdefinestyle{mystyle}{
    basicstyle=\ttfamily\footnotesize,
}
\definecolor{electriclime}{rgb}{0.8, 1.0, 0.0}

\begin{document}
\history{Date of publication ?? ??, ????, date of current version ?? ??, ????.}
\doi{??.????/???.????.???}

\title{Q-gen: A Parameterized Quantum Circuit Generator}
\author{\uppercase{Yikai Mao}\authorrefmark{1},
\uppercase{Shaswot Shresthamali}\authorrefmark{1, 2} \IEEEmembership{Member, IEEE}, 
and \uppercase{Masaaki~Kondo}\authorrefmark{1, 3} \IEEEmembership{Member, IEEE}}
\address[1]{Graduate School of Science and Technology, Keio University, Yokohama, Kanagawa 223-8522, Japan.}
\address[2]{Graduate School of Information Science and Electrical Engineering, Kyushu University, Nishi-ku, Fukuoka 819-0395, Japan.}
\address[3]{RIKEN Center for Computational Science, Kobe, Hyogo 650-0047, Japan.}
\tfootnote{This work is supported by JST COI-NEXT (JPMJPF2221), JST SPRING (JPMJSP2123), and JSPS KAKENHI (JP24K20843).}

\markboth
{Mao, Shresthamali, and Kondo: Q-GEN: A PARAMETERIZED QUANTUM CIRCUIT GENERATOR}
{Mao, Shresthamali, and Kondo: Q-GEN: A PARAMETERIZED QUANTUM CIRCUIT GENERATOR}

\corresp{Corresponding author: Yikai Mao (email: ykmao@acsl.ics.keio.ac.jp)}

\begin{abstract}
Unlike most classical algorithms that take an input and give the solution directly as an output, quantum algorithms produce a quantum circuit that works as an indirect solution to computationally hard problems. In the full quantum computing workflow, most data processing remains in the classical domain except for running the quantum circuit in the quantum processor. This leaves massive opportunities for classical automation and optimization toward future utilization of quantum computing. We kickstart the first step in this direction by introducing Q-gen, a high-level, parameterized quantum circuit generator incorporating 15 realistic quantum algorithms. Each customized generation function comes with algorithm-specific parameters beyond the number of qubits, providing a large generation volume with high circuit variability. To demonstrate the functionality of Q-gen, we organize the algorithms into 5 hierarchical systems and generate a quantum circuit dataset accompanied by their measurement histograms and state vectors. This dataset enables researchers to statistically analyze the structure, complexity, and performance of large-scale quantum circuits, or quickly train novel machine learning models without worrying about the exponentially growing simulation time. Q-gen is an open-source and multipurpose project that serves as the entrance for users with a classical computer science background to dive into the world of quantum computing.
\end{abstract}


\titlepgskip=-15pt

\maketitle


\section{Introduction}

\IEEEPARstart{T}{he} development of quantum mechanics has promoted the birth of quantum computing. To simulate large quantum systems, a new type of computer that operates based on the rules of quantum mechanics will inherently perform better than any classical computer \cite{feynman1982}. In the past few decades, the potential of quantum computing has been demonstrated in many areas including condensed-matter physics, high-energy physics, and chemistry \cite{1308.6253}. Recently, the rapid progress in quantum information theory has revealed that quantum computing can be applied to a much broader field other than just performing simulations. One of the most promising directions is using quantum computing to efficiently solve classically intractable problems \cite{mandi}. 
This is an exciting new area for many computer science researchers. However, they still face a steep learning curve, mostly because the principles used for quantum computation are fundamentally different compared to classical computing. 

The introduction of quantum computing to the classical computing community has been disruptive. The vast majority of literature on quantum computing comes from either mathematics or physics background, 
and practical applications from quantum computers are still not feasible due to heavy noise and limited qubit count. Although quantum computing research is still mostly theoretical, the recent development of NISQ computing has proved that classical computing has great potential to help utilize quantum computers in their current state. Even in the post-NISQ era when large-scale fault-tolerant quantum processors are realized, the probabilistic nature of quantum measurements and the pre/post-processing involved in most quantum algorithms will still require support from classical computers \cite{nisq}.

\Figure[t](topskip=0pt, botskip=0pt, midskip=0pt)[width=0.999\columnwidth]{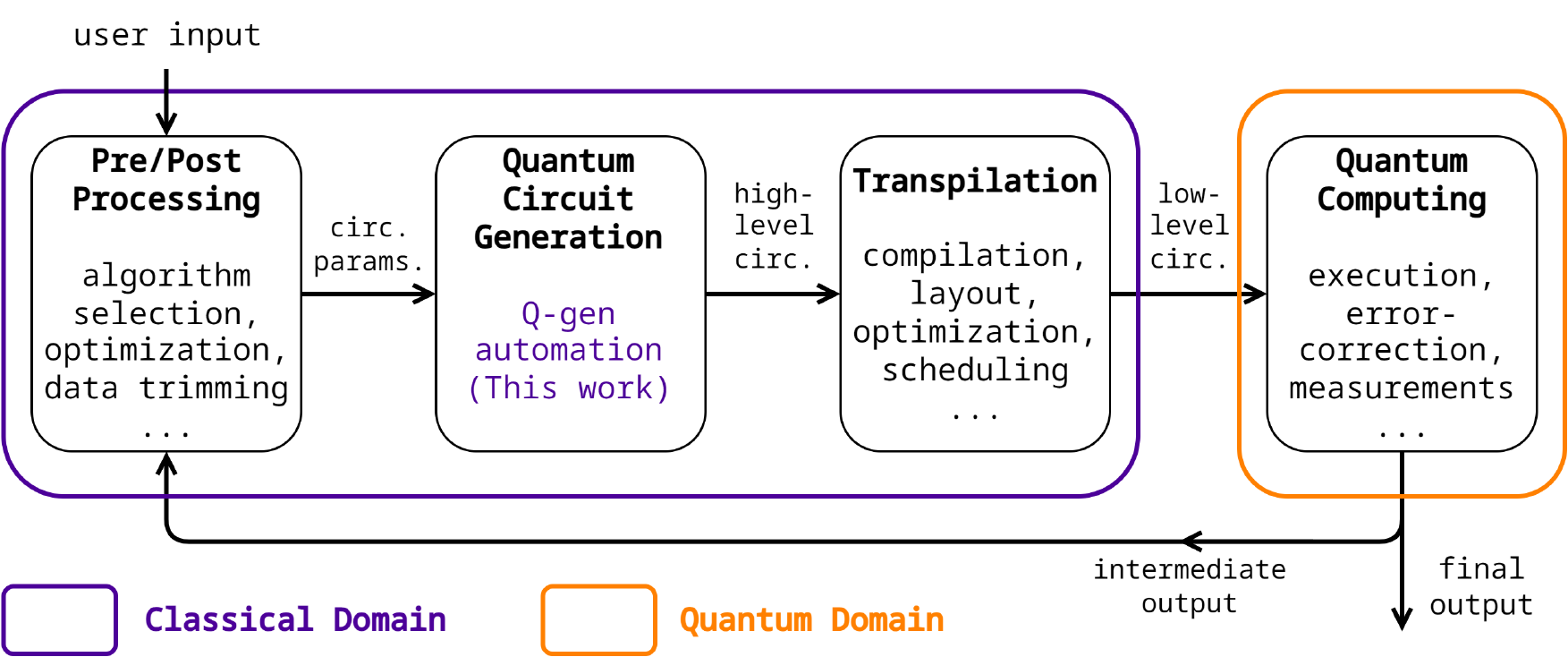}
{A typical workflow of circuit model quantum computing. The classical system takes the computing problem and data from the user to construct a quantum circuit. The circuit is then transpiled into low-level representation to be executed on the quantum hardware. This work is aimed at automating the quantum circuit generation step.\label{workflow}}


Figure \ref{workflow} illustrates a typical circuit model quantum computing workflow, which shows that the majority of computing tasks are performed in the classical domain. We envision that a complete quantum computing process will always involve some classical components, so classical design methods and computing tools are indispensable in accelerating the development of quantum computing. For example, while the current quantum hardware is still catching up on reducing noise, we can improve the high-level circuit to use fewer gates or find better transpilation logic to minimize the qubit count. We can also investigate different routing/placement policies to more efficiently utilize the existing qubit connections, and deploy error-mitigation algorithms to further reduce error rate. 
All these research topics call for a large supply of practical quantum circuits to help with testing and evaluation. However, quickly obtaining various quantum circuits is still not a straightforward task.

This work introduces Q-gen, a quick and efficient tool to create high-level, realistic quantum circuits. For computer science researchers with a limited background in quantum physics, Q-gen removes the knowledge barrier and enables a rapid workflow from problem description to the resultant quantum circuit. Q-gen includes 15 well-known quantum algorithms targeted for different computing problems and offers a modular, parameterized generation process. In addition to specifying the number of qubits of the circuit, Q-gen also provides algorithm-specific generation parameters for more variability, making it possible to produce many circuits with different structures that belong to the same algorithm. The circuits are generated as Qiskit \cite{qiskit} circuit objects, which can be modified and investigated using the native tools from the Qiskit library. This high-level representation offers great portability as it can be either saved directly as a \texttt{.qpy} file or easily translated into other lower-level descriptions like openQASM \cite{openqasm}.

To demonstrate the generation capability of Q-gen, We present the Q-gen quantum circuit dataset consisting of 454 circuits. These circuits are produced with various generation parameters from all 15 available algorithms, ranging from shallow circuits with no \texttt{CNOT} gates to deep circuits with more than 50,000 \texttt{CNOT} gates. Many research projects involve analyzing the output of a quantum circuit, but simulating large circuits can easily cost hours to days. To remove this time overhead for researchers, we also provide ready-to-use simulation results in our dataset as measurement counts and state vectors. Based on the statistics of the Q-gen circuit dataset and the analysis of the simulation result, we organized the quantum algorithms into 5 hierarchical systems with intuitive algorithm complexity ratings, offering a systematic way to quickly understand the origin and application of quantum algorithms.

The contributions of this work include: 
\begin{itemize}
\item A quantum circuit generator based on Qiskit that supports 15 practical algorithms, offering high variability with algorithm-specific generation parameters for quantum algorithm developers.
\item A high-level quantum circuit dataset containing large-scale circuits and their noise-free outputs, provided as measurement counts and state vectors to support the design and optimization of quantum circuits.
\item An organized quantum algorithm system explaining their origin and connections, offering a systematic understanding for new quantum computing researchers.
\item A heuristic analysis of the Q-gen quantum circuit dataset, presents the idea of algorithm complexity categorized by their applications.
\item Open-sourced publication for better community collaboration and future upgrades.
\end{itemize}

The remainder of the paper is structured as follows: Section \ref{background} gives the background and prior research related to quantum circuit generation and optimization. Section \ref{algorithm} introduces all the available algorithms in Q-gen and explains the generation parameters. Section \ref{dataset} explains the design philosophy of the Q-gen algorithm system and presents the analysis of the Q-gen circuit dataset. 
We discuss the practical applications and future improvements of Q-gen in Section \ref{discussion}, and we summarize this work in Section \ref{conclusion}.


\section{Background \& related works} \label{background}

\subsection{Circuit Model Quantum Computing} \label{background1}
The quantum circuit computing model is analogous to classical digital computing, where computations are performed using a sequence of logic gates. In the quantum circuit model, quantum computations are carried out using quantum gates, which are a series of unitary matrix operators that introduce a state or phase transition on the qubit(s) it acts on. Qubits are the fundamental units of quantum information, capable of existing in a superposition of states and entangling with others. These properties enable quantum computers to solve certain problems exponentially faster than classical computers.

In a typical circuit model quantum computing workflow, the quantum algorithm is described using a high-level quantum circuit, and then circuit transpilation compiles the circuit into low-level hardware-specific instructions \cite{1604.01401} for execution. In this process, the circuits are optimized to reduce extra SWAP gates \cite{2111.04572, 2212.05666}, and various policies \cite{2209.15512, 1805.10224, 1809.02573} are applied to search for the best placement of the qubits on the quantum processor. Additionally, software error mitigation \cite{2210.00921, 2209.06864} and hardware error correction \cite{2103.14209, 9773181} techniques are developed to reduce the negative effect of decoherence and communication noise. 



\subsection{Quantum Circuit Generation and Dataset}

Several open-source frameworks exist to help automate or assist in creating quantum circuits. Qiskit \cite{qiskit}, Cirq \cite{cirq}, and Braket \cite{braket} are three major quantum computing frameworks that provide pre-built components and templates for generating higher-level quantum algorithms. All the frameworks also support the OpenQASM \cite{openqasm} representation for lower-level representation of quantum circuits.

Currently, we believe that no projects are specifically targeted for high-level quantum circuit generation, but many works focused on quantum benchmarking usually include some low-level circuit dataset or a lightweight circuit generator with limited functionality. The SupermarQ \cite{2202.11045} benchmark suite includes 8 application circuits and a circuit generator parameterized by the number of qubits. The QASMBench \cite{2005.13018} benchmark suite provides more than 50 openQASM implementations of various scales of quantum circuits in different qubit settings. The Application-Oriented Benchmarks \cite{2110.03137} offer 11 algorithms categorized into 3 groups with a specialized generator for creating random benchmark circuits.

\subsection{AI for Quantum Computing}

Artificial Intelligence (AI) has experienced explosive development since the mid-2010s. In addition to the classical optimization methods mentioned in Section \ref{background1}, there is a growing interest in using AI to help accelerate the development of quantum computing. This also gives a strong motivation for our work because machine learning models require a large number of quantum circuits with high variability for efficient training.

For ML-based circuit optimization, there are works on using Large Language Models (LLM) to design better quantum circuits \cite{2307.08191} and using Transformers to simplify \texttt{CNOT} circuits \cite{teachingzx}. Deep reinforcement learning techniques are also explored to minimize the number of \texttt{T} gates \cite{2402.14396}. Many works proposed different machine learning models for predicting the output fidelity of quantum circuits \cite{2102.02369, 2212.00677, 2303.17523, 2210.16724, 2404.06535}, which can be used for finding better layout and improve \texttt{CNOT} routing. These works amplify the demand for a more realistic and flexible quantum circuit dataset for better training. New research areas that have the potential for AI integration can also benefit from a large supply of quantum circuits. One example is circuit knitting, a technique of cutting the full circuits into smaller pieces for more efficient execution \cite{2309.07857, 2205.00016, 2303.10788}.


\Figure[t](topskip=0pt, botskip=0pt, midskip=0pt)[width=1\textwidth]{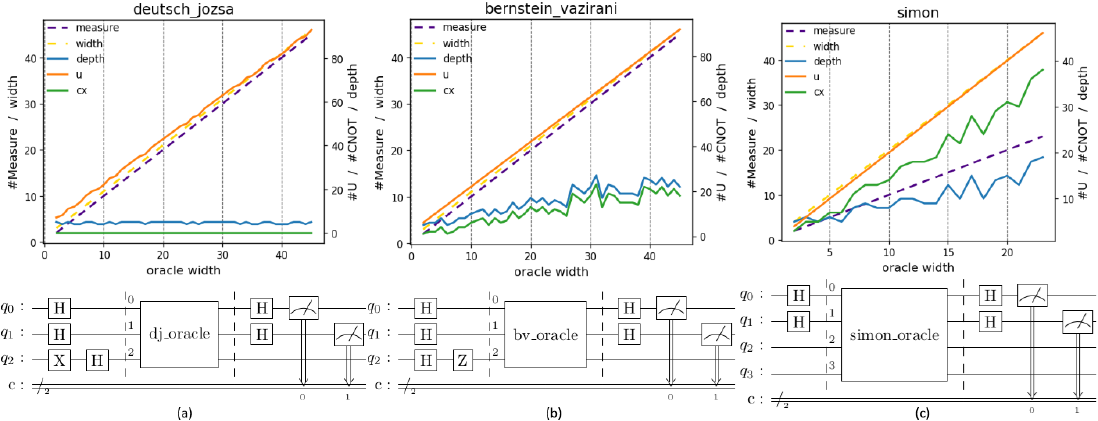}
{Q-gen quantum circuit dataset statistics and example circuit of the (a) Deutsch-Jozsa algorithm, the (b) Bernstein-Vazirani algorithm, and (c) Simon's algorithm. Circuit width/depth is the number of qubits/steps in the circuit. U/CX/Measure means Unitary (single-qubit) gate/CNOT (two-qubit) gate/Measurement gate. For this group of algorithms, the main generation parameter (\texttt{problem size}) is the qubit width of the oracle. The gate and circuit statistics all increase linearly, which indicates that these circuits are relatively simple to generate and simulate. Notice that there are no CNOT gates in the Deutsch-Jozsa algorithm. The circuit visualizations in this algorithm group indicate a clear initialization-oracle-measurement structure. \label{alg_group_0}}

\section{Q-gen: Algorithm Design \& Analysis} \label{algorithm}

In this Section, we introduce the quantum algorithms included in Q-gen. We explain the significance of the algorithms, the Q-gen implementation with the available generation parameters, and the potential applications. Each algorithm is accompanied by a circuit complexity visualization consisting of gate statistics and circuit size data, controlled by the algorithm-specific \texttt{problem size} parameter. The example circuits included in Figure \ref{alg_group_0} to Figure \ref{alg_group_4} are generated by Q-gen using the simplest available generation parameters.


\subsection{Deutsch-Jozsa Algorithm}

The Deutsch-Jozsa algorithm \cite{dj} determines whether a special boolean function \(f\) is constant or balanced: a constant \(f\) will always output all 0 or 1 regardless of the input, and a balanced \(f\) will always output 0 for half of the input and 1 for the other half. Classically, at least two queries to \(f\) are required to find the answer, but the quantum version of this algorithm only requires one query.

Q-gen can generate the oracle with both types of boolean functions, and the oracle width is decided by the \texttt{problem size} parameter. The constant oracle is a very simple oracle that only applies \texttt{X} gate to the output qubit, the balanced oracle adds more circuit complexity by creating entanglement with \texttt{CNOT} gates. Intuitively, as the \texttt{problem size} grows, the depth of the generated circuit remains constant for the constant oracle, and the number of single-qubit gates grows linearly, shown in Figure \ref{alg_group_0}. If the oracle is balanced, the depth and number of \texttt{CNOT} gates both grow linearly. 

This algorithm is useful for generating simple circuits (no \texttt{CNOT} gates) with constant depth but growing width. The only other algorithm in Q-gen that has a similar circuit complexity pattern is the quantum key distribution algorithm, but its number of single-qubit gates and measurements is doubled.

\subsection{Bernstein-Vazirani Algorithm}

The Bernstein-Vazirani algorithm \cite{bv} solves problems similar to but harder than the Deutsch-Jozsa problem. The hidden function in this algorithm returns 0 or 1 based on the bitwise product of the input string with a hidden ``secret'' string \(s\). The goal is to find \(s\) with as few queries to the oracle as possible. The best classical solution must query every input bit of \(f\). In contrast, the quantum version of this algorithm only requires one query.

Q-gen can generate the Bernstein-Vazirani oracle randomly or based on a binary string, with its width controlled by \texttt{problem size}. The oracle places the \texttt{control} of the \texttt{CNOT} gate on the input qubit if \(s=1\) and leaves the qubit untouched if \(s=0\), the \texttt{target} of all \texttt{CNOT} gates are placed on the additional phase-kickback qubit. The growth pattern of the Bernstein-Vazirani algorithm is similar to the Deutsch-Jozsa algorithm, but as \texttt{problem size} increases, the depth and number of \texttt{CNOT} gates all grow linearly at a higher rate. 

This algorithm is good for generating elementary quantum circuits with some entanglement. Also, the output of the Bernstein-Vazirani circuit is easily verifiable without any post-processing, which makes it a good candidate for quantum benchmarking.

\subsection{Simon's Algorithm}


Simon's algorithm \cite{simon} has proved that quantum computers can offer exponential speed-up over their classical counterparts. The oracle function \(f\) in Simon's problem can be one-to-one or two-to-one according to a secret string \(s\). Similar to the Deutsch-Jozsa algorithm and the Bernstein-Vazirani algorithm, we want to determine the type of \(f\) as fast as possible. While the best classical solution requires \(O(\sqrt{2^{n}})\) queries, the quantum algorithm only requires \(O(n)\) queries \cite{1610.01920}, achieving exponential speed-up. 

The circuit structure of Simon's algorithm is special in terms of circuit width. Although \texttt{problem size} is still defined as the oracle width, the generated circuit will have a qubit count that doubles the number of \texttt{problem size}. Upon measurements, the result will be multiple guesses related to the secret string \(s\), and \(s\) can be revealed with some moderate post-processing. Simon's algorithm requires more \texttt{CNOT} gates than the other two query algorithms in Q-gen, and the growth rate for the number of \texttt{CNOT} gates is approximately 2$\times$ of circuit depth, as shown in Figure \ref{alg_group_0}.

Simon's algorithm has the highest circuit complexity compared with the Deutsch-Jozsa algorithm and the Bernstein-Vazirani algorithm, considering their simulation time and the number of \texttt{CNOT} gates at the same \texttt{problem size}. Since it requires multiple shots and post-processing to find the correct answer, Simon's algorithm is good for testing the full quantum\(+\)classical hybrid computing workflow.

\subsection{Quantum Fourier Transform}

Quantum Fourier Transform (QFT) \cite{qft} is the quantum implementation of the classical Discrete Fourier Transform (DFT). To perform DFT on \(2^n\) elements, classical algorithms like the Fast Fourier Transform (FFT) require \(O(n2^n)\) operations. QFT only needs \(O(n^2)\) operations, which is exponentially fewer than the classical algorithms \cite{mandi}. Note that QFT does not directly solve any specific problems, if the qubits are measured in the computational basis, the result will appear random as they have been transferred to the Fourier basis.

Q-gen takes \texttt{problem size} as the circuit width and generates the barebone QFT subroutine. Although this circuit can be directly integrated into other higher-level algorithms, it is not directly verifiable. Q-gen provides three generation parameters that make verification possible: the qubits can be \texttt{initialized} in the Fourier basis using an integer input, and the QFT subroutine can be \texttt{inverted} to transform the qubits into the computational basis. By \texttt{measuring} the qubits, the result will be the integer number used for initialization. While QFT's circuit complexity growth pattern is exponential, as shown in Figure \ref{alg_group_1}, it is still a very efficient implementation since the depth only reaches around 4,000 when \texttt{problem size} is 47 qubits.

QFT is the building block for many practical quantum algorithms, so reducing its circuit complexity and at the same time increasing its accuracy is an important research topic. For instance, the approximate QFT algorithm can ignore phase rotations below a certain threshold and still yield the correct result with acceptable error \cite{1803.04933}. Q-gen's parameterized generation offers a quantitative approach to quickly analyze the behavior of various QFT circuits.

\Figure[t](topskip=0pt, botskip=0pt, midskip=0pt)[width=1\textwidth]{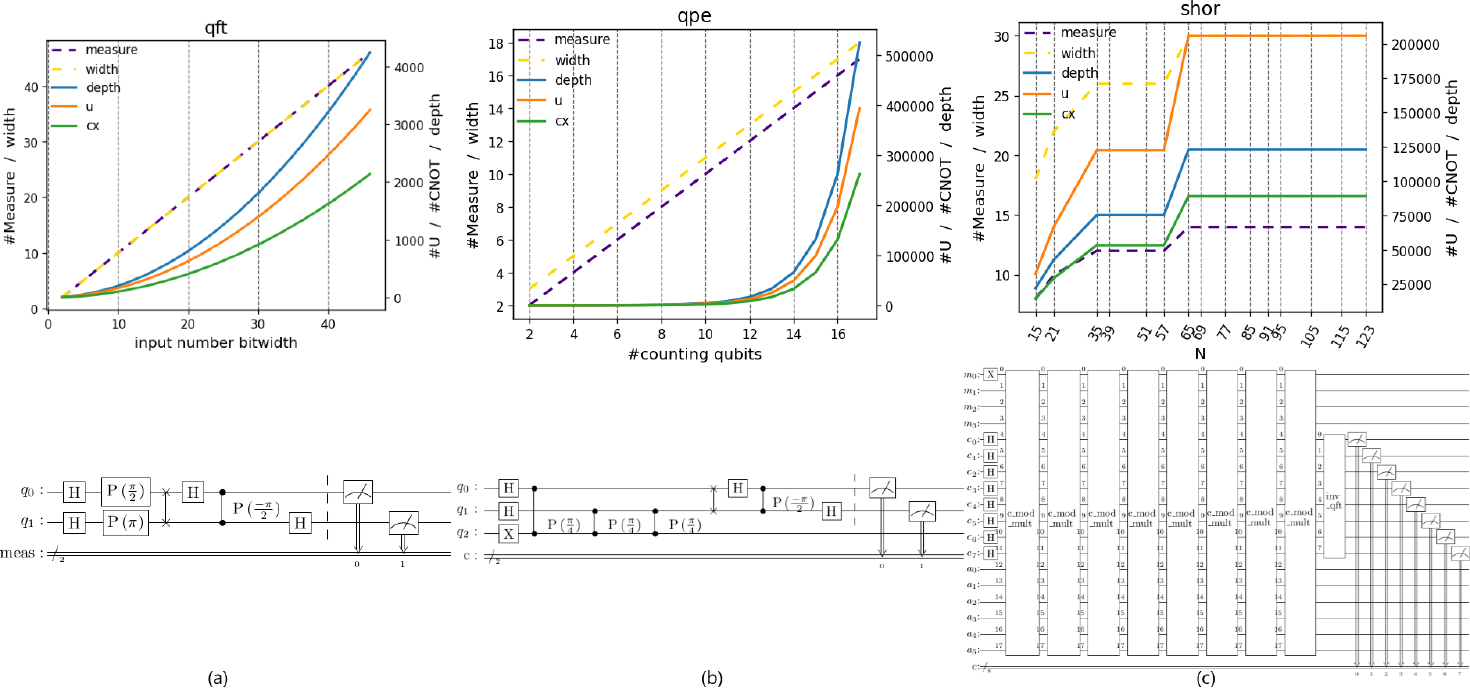}
{Q-gen quantum circuit dataset statistics and example circuit of (a) quantum Fourier transform (QFT), (b) quantum phase estimation (QPE), and (c) Shor's algorithm. Circuit width/depth is the number of qubits/steps in the circuit. U/CX/Measure means Unitary (single-qubit) gate/CNOT (two-qubit) gate/Measurement gate. For QFT, \texttt{problem size} is the classical bitwidth of the input number to be transformed. For QPE, the circuits are generated by the number of counting qubits. For Shor's algorithm, N is the number to be factored. This group of algorithms shows a clear exponential growth pattern with a single-qubit gate dominance over the two-qubit gates, and the circuit visualizations show heavy CNOT connection patterns. High-resolution circuit visualization is hosted on the Q-gen dataset Wiki page.\cite{github_wiki} \label{alg_group_1}}

\subsection{Quantum Phase Estimation}

Quantum phase estimation (QPE) \cite{qpe} is one of the key applications of QFT. This algorithm uses phase-kickback and inverse-QFT to estimate the phase/eigenvalue of any unitary operator \(U=e^{2\pi i \theta}\). The phase angle \(\theta\) is recorded in the counting qubits by the iterative \texttt{controlled-U} gates applied on the eigenstate qubit. With \(n\) counting qubits and the measurement result equal to \(x\), \(\theta\) can be estimated as \(\theta \approx x/2^n\).

The QPE circuits from Q-gen have one eigenstate qubit for \(U\) to act on, and \(n\) counting qubits, controlled by \texttt{problem size}. 
The estimation precision is set by the number of counting qubits \(n\), with the resolution equal to \(1/2^n\).
Due to the high implementation cost of the iterative \texttt{controlled-U} gates, the complexity of QPE circuits grows exponentially, reaching a depth more than \(5.0\times10^5\) at \texttt{problem size} = 17. This causes other QPE-based algorithms to have an even higher gate count, like the quantum walk search algorithm.

QPE has many practical applications because some classically hard problems can be condensed to phase estimation, the most famous examples include period-finding and prime factorization. However, it is still hard to implement QPE on current NISQ hardware due to its high circuit complexity. Although we can reduce the number of counting qubits, it also causes the estimation accuracy to drop. Q-gen can generate a large sample of QFT circuits to form a baseline of estimation accuracy, which can then be used to study the trade-off between reduced gate count and loss of accuracy.

\Figure[t](topskip=0pt, botskip=0pt, midskip=0pt)[width=1\textwidth]{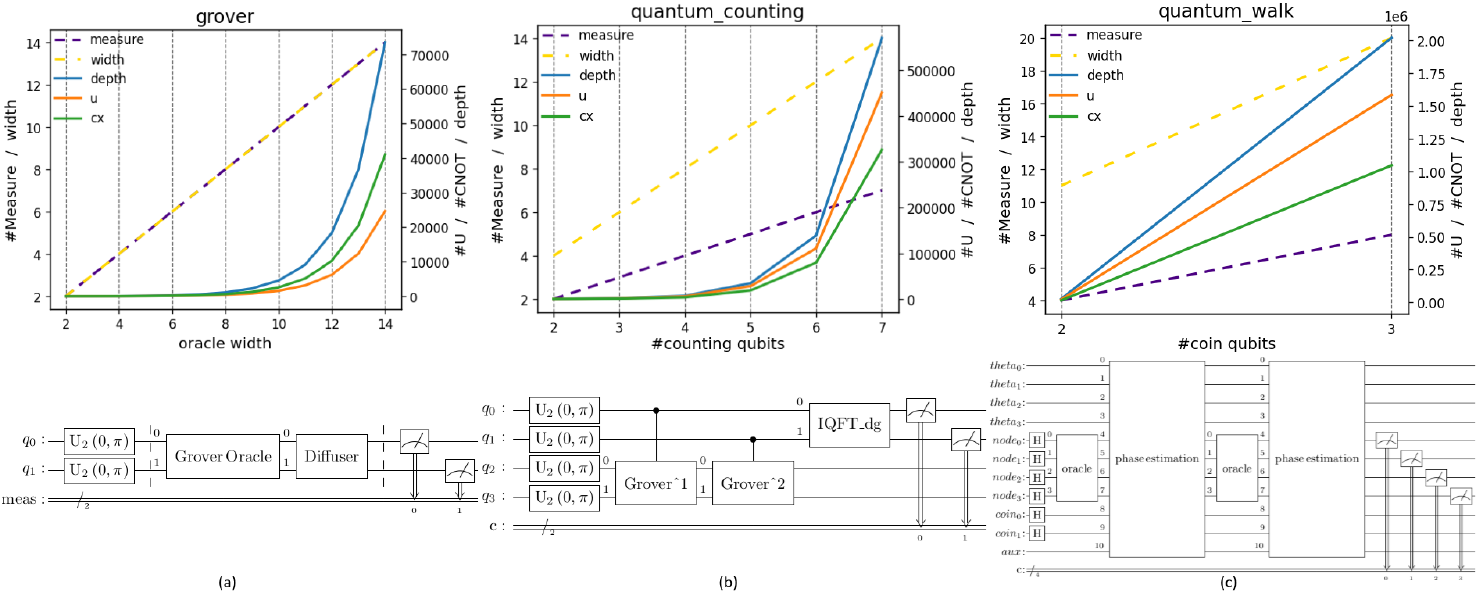}
{Q-gen quantum circuit dataset statistics and example circuit of (a) Grover's algorithm, (b) quantum counting, and (c) quantum walk algorithm. Circuit width/depth is the number of qubits/steps in the circuit. U/CX/Measure means Unitary (single-qubit) gate/CNOT (two-qubit) gate/Measurement gate. For Grover's algorithm, \texttt{problem size} is the number of qubits of the Grover oracle. For quantum counting, the main generation parameter is the number of counting qubits. For quantum walk, the main generation parameter is the number of coin qubits. These three algorithms solve practical problems but they require very high quantum resources, notably in quantum walk algorithm the number of CNOT gates is over 1,000,000 when there are only 3 coin qubits. The circuit visualizations show that all of the algorithms in this group utilize the Grover's oracle. High-resolution circuit visualization is hosted on the Q-gen dataset Wiki page.\cite{github_wiki} \label{alg_group_2}}

\subsection{Shor's Algorithm}
Shor's algorithm \cite{shor} can factor any large number \(N\) in polynomial time, better than every known classical algorithm. Under the hood, the core problem that contributes to this quantum speedup is the period-finding problem. If the period \(r\) of the modular exponential function \(f(x)=a^x mod N\) can be found in polynomial time, the factor(s) of \(N\) can also be found in polynomial time. Shor's algorithm uses QPE to accelerate the period-finding process, the subsequent search for factor(s) can be done efficiently using classical algorithms.

Shor's algorithm is a quantum-classical hybrid algorithm that utilizes classical pre-processing to return a result before the QPE subroutine if \(N\) is simple. To ensure the algorithm reaches the quantum part, \(N\) has to be reasonably hard to factor. Specifically for Shor's algorithm, \(N\) has to be odd and not formed by \(m^n\) for \(m \ge 1\) and \(n \ge 2 \). Additionally, a random guess of \(a\) is required to kick-start the modular exponential function, and \(a\) must be a coprime of \(N\) \cite{mandi}. Q-gen automatically generates the suitable \(a\) and provides a list of \(N\) (as \texttt{problem size}, up to 123) ready to use for the quantum circuit. The main difficulty in Shor's algorithm is the quantum implementation of the modular exponential function, many customized implementations exist focusing on minimizing the gate count or using a specialized basis gate set \cite{1903.00768
, 9806084}. Q-gen implements a relatively efficient modular exponential circuit using \(2n+3\) qubits with \(n\) equal to the bit-length of \(N\) \cite{0205095, shor_github}. Most importantly, this implementation is general regardless of \(N\), ideal for Q-gen's parameterized generation process.

Like most quantum algorithms, Shor's algorithm only displays quantum advantage over classical algorithms when the problem size is sufficiently large \cite{2307.00523}. Although the ability to factor large prime numbers in polynomial time potentially breaks the RSA cryptosystem, it is still unclear whether this can be realized on a practical quantum computer in the foreseeable future. As of 2023, the largest factored RSA number has 829 bits (RSA-250) \cite{rsa}. A quick analysis using Q-gen shows that factoring the same number using a quantum computer requires at least 1,661 qubits, excluding the auxiliary qubits.

\subsection{Grover's Algorithm}

Grover's algorithm \cite{grover} uses the Grover oracle \(G\) and Diffusion operator \(D\) to accelerate search problems.
\(G\) contains logical computations that mark the target states with a negative phase, and \(D\) then amplifies the measurement probability of the marked states while simultaneously reducing the probability of other states. For an unstructured database with \(N\) items, Grover's algorithm can find all \(M\) targets after approximately \(\sqrt{N/M}\) Grover iterations (\(G+D\)), providing a quadratic speedup over the classical algorithms.

Q-gen can automatically generate \(G\) based on \texttt{problem size} with the optimal \(M\) that only requires 1 Grover iteration. The target state in \(G\) can also be specified as an input string. Q-gen will append the appropriate number of Grover iterations to the circuit, or take this number as another input. The width of Grover's circuit always equals the number of measurement gates and they both grow linearly, but its circuit complexity grows exponentially as the search space increases. Noticeably, compared with the quantum counting algorithm, the number of \texttt{CNOT} gates grows faster than the number of single-qubit gates, as shown in Figure \mbox{\ref{alg_group_2}.}

Q-gen offers a quick way to statistically and empirically analyze the performance of Grover's algorithm, especially when the search space is large. The parameterized circuit generation also enables an in-depth investigation of the algorithm. For example, testing how different numbers of Grover iterations affect result accuracy.

\Figure[t](topskip=0pt, botskip=0pt, midskip=0pt)[width=1\textwidth]{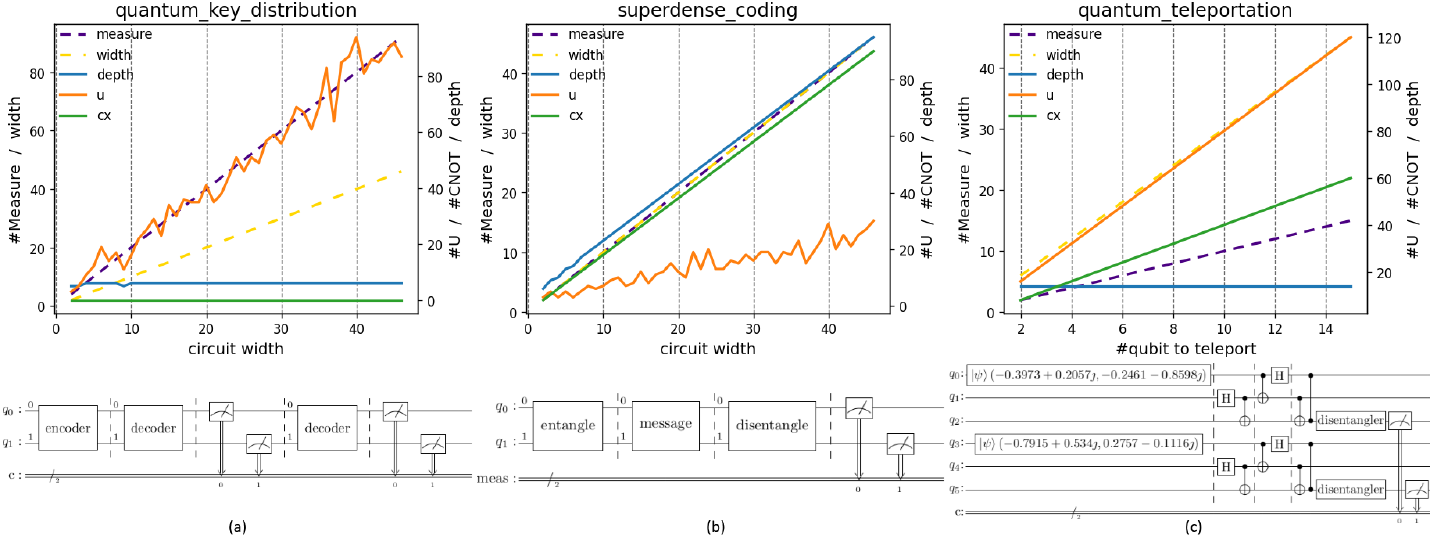}
{Q-gen quantum circuit dataset statistics and example circuit of (a) quantum key distribution, (b) superdense coding, and (c) quantum teleportation. Circuit width/depth is the number of qubits/steps in the circuit. U/CX/Measure means Unitary (single-qubit) gate/CNOT (two-qubit) gate/Measurement gate. For both quantum key distribution and superdense coding, the \texttt{problem size} is the width of the circuit. In quantum teleportation, the main generation parameter is the number of qubits to be teleported. This is another group of simple algorithms that shows linear growth, notice that there are no CNOT gates in the quantum key distribution algorithm. The circuit visualizations demonstrate a clear modular pattern, which is easy for parallel expansion.\label{alg_group_3}}

\subsection{Quantum Counting}

The quantum counting algorithm \cite{counting} can estimate the number of solutions \(M\) inside \(N\) items, it is a combination of Grover's algorithm and quantum Fourier transformation (QFT). This algorithm can be seen as a prerequisite for Grover's algorithm since Grover's algorithm requires \(M\) to calculate the correct number of iterations. It can also determine whether a solution even exists inside \(N\) items, which can accelerate certain NP-complete problems like the Hamiltonian Cycle problem \cite{mandi}.

There are two groups of qubits inside the quantum counting algorithm: the \textit{searching qubits} iterate over the search space \(N\), and they use controlled Grover's operator to mark \(M\) on the \textit{counting qubits}. The default setting of Q-gen takes \texttt{problem size} as the number of counting qubits and assigns an equal number of searching qubits to ensure optimal search results. The number of solutions \(M\) can be randomly generated or specified. Under the default setting, as \texttt{problem size} increases, the width of the generated circuits grows faster than the number of measurement gates. Compared with Grover's algorithm, The number of single-qubit gates grows faster than the number of \texttt{CNOT} gates.

Q-gen's generation settings can be easily changed to investigate the efficiency and accuracy of quantum counting under different problem spaces. More importantly, this algorithm is another great candidate for benchmarking the quantum\(+\)classical hybrid computing workflow, since it contains two famous quantum subroutines: Grover's operator and inverse-QFT, plus moderate classical post-processing.

\subsection{Quantum Walk Algorithm}

Quantum walk \cite{walk} is the quantum version of the random walk search algorithm. On each walking step, the ``walker'' enters superposition to search all the nodes on a graph simultaneously. For a graph with \(N\) nodes and \(M\) targets, approximately \(1/\sqrt{|M|/N}\) iterations are required to find all targets \cite{1907.09415}. This is another quantum search algorithm that provides a quadratic speedup.

Q-gen implements the coined quantum walk, which uses a coin to direct how the ``walker'' moves. 
Specifically, Q-gen uses the Grover coin, equivalent to the Diffusion operator \(D\) from Grover's algorithm. Each iteration contains a phase oracle that marks the target states, followed by quantum phase estimation using the coined walk steps. Q-gen takes \texttt{problem size} as the width of the coin, then automatically determines the width of the node qubits, and the number of counting qubits for phase estimation. Due to the extremely high gate cost of the coined walk steps, under the default setting, a 3-qubit coin will generate a circuit with more than \(1.0\times10^6\) \texttt{CNOT} gates and more than \(1.5\times10^6\) single-qubit gates.

Although this is a resource-intensive algorithm, many parameters are still available in Q-gen to simplify its circuit structure. The number of counting qubits for phase estimation can be reduced, or the width of the node qubits can be set smaller. The number of iterations can also be specified, potentially lowering the search accuracy.

\subsection{Quantum Key Distribution}

Quantum key distribution \cite{key} is a secure quantum communication protocol based on one primary quantum principle: measurement collapses superposition. Suppose a sender prepares a qubit in a specific basis and sends it through a quantum communication channel. In that case, the receiver can retrieve the same information by measuring the qubit in the same basis. However, if the qubit is measured in a different basis before reaching the receiver, the qubit collapses prematurely and the receiver will get a random result.

Q-gen generates the full quantum circuit simulating the key distribution process. The \texttt{problem size} controls the qubit width of the circuit, equivalent to the bit width of the key. The \texttt{interception} parameter controls whether an ``attacker'' will be inserted before the final measurement. This parameter causes the circuit to have mid-circuit measurements, which is the only occurrence among Q-gen's algorithms. Q-gen can also simulate the circuit and output the information measured at each stage, including if the ``attacker'' has been detected. The generated circuits only contain single-qubit gates with a linear growth pattern. The depth has two fixed variations depending on whether \texttt{interception} is true.

This key distribution protocol is not designed to be risk-free. For example, if the ``attacker'' inadvertently picked the same measurement basis as the sender, the interception will go undetected. However, adding more qubits to the key can significantly lower this risk. Q-gen's generation parameters provide a quick method to analyze the quantum circuit's complexity under different security conditions.

\Figure[t](topskip=0pt, botskip=0pt, midskip=0pt)[width=1\textwidth]{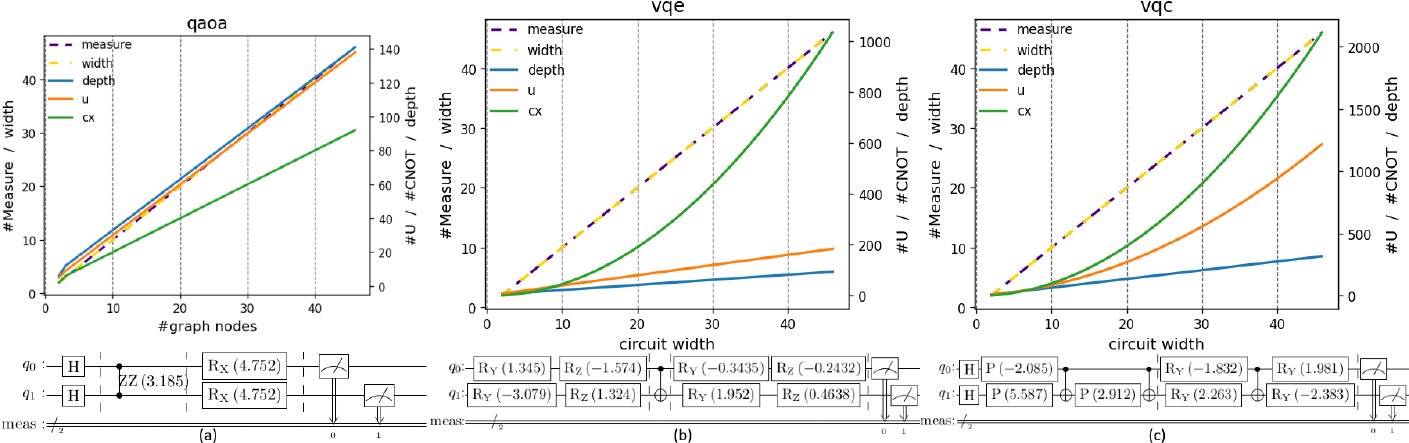} %
{Q-gen quantum circuit dataset statistics and example circuit of the (a) quantum approximate optimization algorithm (QAOA), 
the (b) variational quantum eigensolver (VQE), and the (c) variational quantum classifier (VQC). Circuit width/depth is the number of qubits/steps in the circuit. U/CX/Measure means Unitary (single-qubit) gate/CNOT (two-qubit) gate/Measurement gate. For QAOA, the \texttt{problem size} is the number of nodes in the input graph. For both VQE and VQC, the main generation parameter is the width of the circuit. These algorithms have high variability which expresses linear and exponential growth patterns depending on the generation parameters. The circuit visualizations indicate that the variational circuit depends heavily on arbitrary rotation gates. \label{alg_group_4}}

\subsection{Superdense Coding}

Superdense coding \cite{superdense} is a quantum communication protocol that encodes 2 bits of classical information using 1 qubit. This protocol requires the sender and receiver to share a pair of entangled qubits before the transmission starts, usually prepared by a third party. The sender can pick from 4 quantum gates to apply on the qubit, corresponding to 4 possible binary numbers using 2 classical bits. After the receiver obtains this qubit, the message can be decoded by disentangling the qubit pair. Superdense coding is also secure, as the entanglement will be destroyed if any qubit is prematurely measured before reaching the receiver.

Q-gen generates the three steps of superdense coding: entangling, encoding, and disentangling. The \texttt{problem size} controls the width of the circuit, equal to the number of qubits used for transmission. The number of \texttt{CNOT} gates grows linearly alongside the depth of the circuit. The number of qubits to encode, and the single-qubit gates used for each encoding, can all be specified or randomly generated. 

The superdense coding circuits are highly structured and easy to scale up. This makes it a great candidate for benchmarking the entanglement ability, and the measurement fidelity of the quantum hardware.

\subsection{Quantum Teleportation}

Quantum teleportation \cite{teleport} recreates the sender's qubit on the receiver's side using a pair of entangled qubits and classical bits. The sender's qubit is processed with one of the qubits from the entangled pair, and then measured to obtain a 2-bit classical data. This data is transferred to the receiver through a classical channel so the receiver can reconstruct the sender's qubit on top of the other qubit from the entangled pair. This protocol can be seen as the opposite of superdense coding.

The minimal teleportation circuit in Q-gen involves 3 qubits. The state to be teleported is initialized on \(q_0\), the entangled pair is created on \(q_1\) and \(q_2\). After the deferred measurement on \(q_0\) and \(q_1\), the original state is recreated on \(q_2\). An inverse-initializer is then placed on \(q_2\), if the state is successfully teleported, it should inverse the qubit back to \(\ket{0}\). The \texttt{problem size} defines how many sets of teleportation circuits to create, thus the generated circuit is analogous to a classical parallel communication bus. The number of \texttt{CNOT} and single-qubit gates follow a linear growth pattern, and the circuit's depth is always fixed, as shown in Figure \ref{alg_group_3}.

The quantum teleportation algorithm in Q-gen is easily verifiable because the ideal measurement should always be all \(\ket{0}\). Additionally, because every set of teleportation circuits is physically isolated, this algorithm can potentially detect crosstalk errors between different regions on a quantum processor.

\subsection{Quantum Approximate Optimization Algorithm}

The quantum approximate optimization algorithm (QAOA) \cite{qaoa} can approximate the solution of the combinatorial optimization problem. For complex optimization problems like MaxCut and Max-kXOR, QAOA gives higher quality solutions and takes less time compared to the classical algorithms \cite{2306.09198}. The cost function from the combinatorial optimization problem is encoded into a quantum circuit called variational form, then the expectation value can be measured repeatedly to optimize the parameters of the cost function, ultimately converging to the solution.

The QAOA circuit in Q-gen is based on the popular MaxCut problem, which tries to partition a graph so that the number of edges between the two sets of nodes is maximum. Q-gen generates the variational form based on the input graph and randomly initializes all the parameters. The number of nodes in the graph is defined by \texttt{problem size}, and each edge in the graph is translated to an \texttt{Rzz} gate which decomposes into two \texttt{CNOT} gates and one \texttt{Rz} Gate. The depth and gate counts of the QAOA circuit grow linearly with \texttt{problem size}. Even if the input graph is fixed and the repetitions of the variational form increase, the circuit's depth still grows linearly, making QAOA a very efficient quantum algorithm \cite{qaoa}.

QAOA is a heuristic algorithm that gives better approximations after every measurement and optimization cycle. Although it has the potential to outperform certain classical algorithms, the implementation of QAOA still heavily depends on the topology of the input graph. Q-gen can help researchers quickly explore the circuit structure of QAOA under different input graphs.

\Figure[t](topskip=0pt, botskip=0pt, midskip=0pt)[width=1\textwidth]{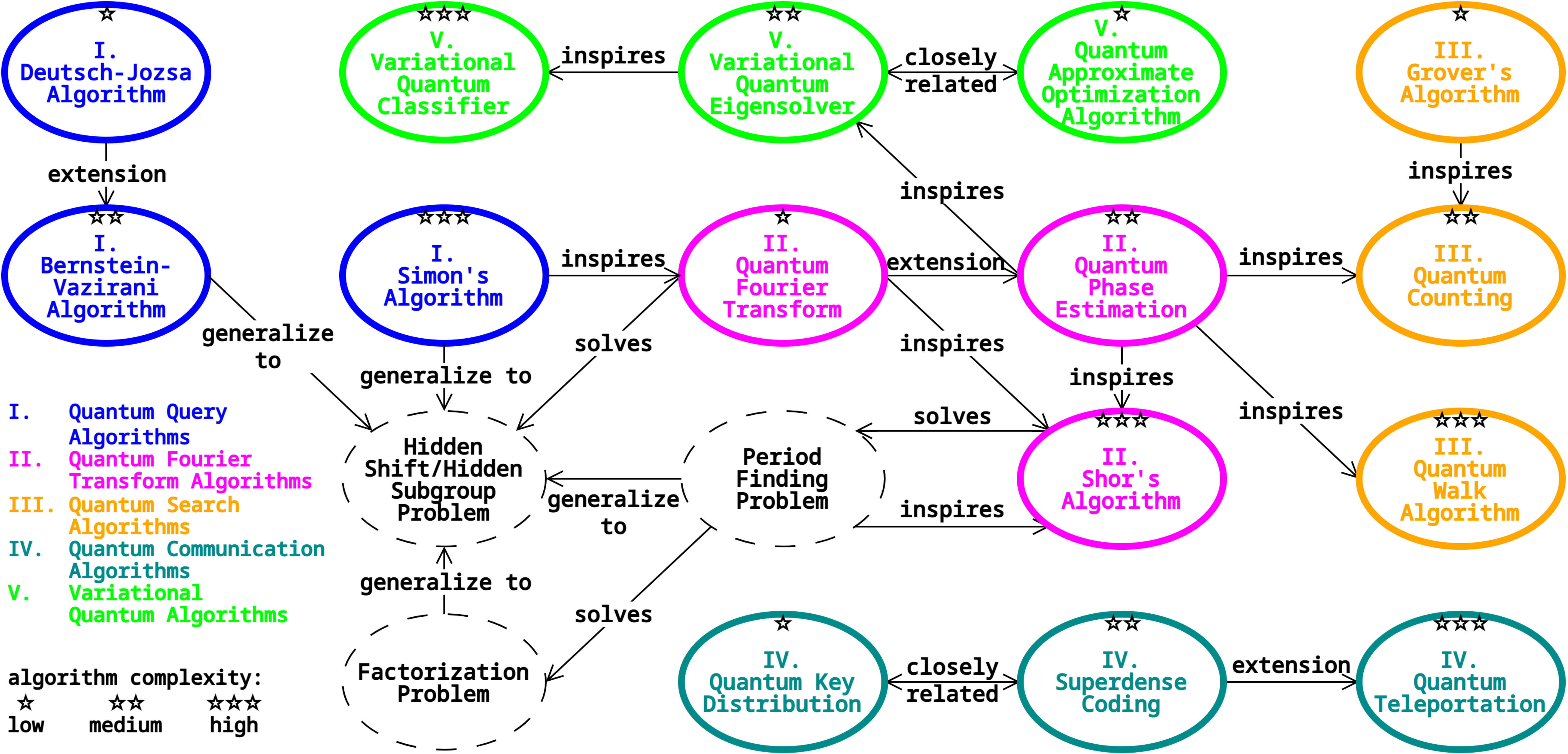}
{The complete Q-gen algorithm system illustration. The algorithm's complexity rating is designed to be compared vertically within its category instead of horizontally across different algorithm categories. To help form the connection between different algorithms, we include some generalized quantum computing problems, indicated by the dashed circles. The connections are gathered from various textbooks, lecture notes, and scientific papers. \cite{mandi} \cite{alg_system_1} \cite{alg_system_2} \cite{vqe} \cite{alg_system_4} \label{alg_system}}

\subsection{Variational Quantum Eigensolver}

The variational quantum eigensolver (VQE) can estimate the minimum eigenvalue of a physical system represented by a matrix. Since the current quantum hardware is not capable of running deep QPE algorithms, VQE is designed as an alternative to efficiently approximate the solution with the help of a classical optimizer \cite{vqe}. VQE has many practical applications in the fields of chemistry and physics. For example, finding the ground state energy of molecules \cite{2103.08505}. 

The implementation of the variational forms for VQE depends on the specific system being simulated. Q-gen fully utilizes Qiskit's circuit library to give a wide variety of generation parameters. The choices of the single-qubits gates for the initial mixing unitary are generated according to the \texttt{gates} parameter, the entanglement pattern (full, linear, etc.) is generated based on the \texttt{entanglement} parameter, and the \texttt{repeat} parameter controls the number of repetitions of the full variational form. Like the other variational algorithms in Q-gen, the rotation angle of the gates can all be specified or randomly generated.

Because the VQE algorithm in Q-gen is highly customizable, it enables explicit control of the percentage of single-qubit and two-qubit gates, which produces circuits with vastly different structures.

\subsection{Variational Quantum Classifier}

The variational quantum classifier (VQC) is the quantum version of the traditional neural network classifier \cite{1611.09347}. By embedding classical data into a quantum feature map circuit, a variational form can be trained on this feature map just like training the classical neural networks. Machine learning is another promising application of variational quantum algorithms. As the datasets and models get larger, quantum computing can potentially make training more efficient by utilizing the exponentially growing parameter space represented by more entangled qubits.

Q-gen implements a common VQC circuit using the \texttt{ZZFeatureMap} and the \texttt{RealAmplitudes} variational form \cite{qiskitmachinelearning}. The repetition number of the feature map and the variational form can be separately controlled, depending on the dataset embedding design and the number of trainable parameters. Additionally, different feature maps provided by the Qiskit library can be easily swapped in, like the \texttt{PauliFeatureMap} and \texttt{ZFeatureMap} \cite{qiskitcircuitlib}. Compared with QAOA and VQE circuits in Q-gen, VQC circuits contain significantly more \texttt{CNOT} gates due to the entanglement introduced by the feature map circuit, as shown in Figure \ref{alg_group_4}.

By taking advantage of the striking similarities between the connected neurons and the entangled qubits, VQC demonstrates how quantum computing can augment classical computing tasks. The circuit generation parameters in Q-gen can be tuned similarly to the traditional configuration variables like learning rate and batch size.


\begin{table*}[!htb]
\caption{Q-gen Dataset Generation Summary}
\label{table_dataset}
\centering
\resizebox{\textwidth}{!}{%
\begin{tabular}{@{}llllllllllll@{}}
\toprule
Q-gen Category & Algorithm & \begin{tabular}[c]{@{}l@{}}Available \\ Parameters\\ (* = \texttt{problem size})\end{tabular} & \begin{tabular}[c]{@{}l@{}}Dataset \\ Parameters\end{tabular} & \begin{tabular}[c]{@{}l@{}}Mean\\ Generation \\ Time (ms)\end{tabular} & \begin{tabular}[c]{@{}l@{}}\#Circuits \\ in Dataset\end{tabular} & \begin{tabular}[c]{@{}l@{}}Circuit \\ Width \\ Range\end{tabular} & \begin{tabular}[c]{@{}l@{}}Circuit Depth \\ Range\end{tabular} & \begin{tabular}[c]{@{}l@{}}\#Single-Qubit\\ Gates\end{tabular} & \#CNOTs & \#Measure \\ \midrule
Quantum & Deutsch- & oracle width* & [2, \(\infty\)] & 1.72 & 44 & [3, 46] & [5, 5] & [7, 93] & N/A & [2, 45] \\
Query & Jozsa \cite{dj} & oracle type & constant oracle &  &  &  &  &  &  &  \\
Algorithms &  & oracle content & random &  &  &  &  &  &  &  \\ \cmidrule(l){2-11} 
 & Bernstein- & oracle width* & [2, \(\infty\)] & 2.17 & 44 & [3, 46] & [5, 22] & [6, 92] & [1, 18] & [2, 45] \\
 & Vazirani \cite{bv} & oracle content & random &  &  &  &  &  &  &  \\ \cmidrule(l){2-11} 
 & Simon's \cite{simon} & oracle width* & [2, \(\infty\)] & 1.46 & 22 & [4, 46] & [5, 19] & [4, 46] & [3, 38] & [2, 23] \\
 &  & oracle content & random &  &  &  &  &  &  &  \\ \midrule
Quantum & QFT \cite{qft} & input number bitwidth* & [2, \(\infty\)] & 555 & 45 & [2, 46] & [13, 4237] & [9, 3243] & [5, 2139] & [2, 46] \\
Fourier &  & input number initialization & random &  &  &  &  &  &  &  \\
Transform &  & inverse? & True &  &  &  &  &  &  &  \\
Algorithms &  & measurement? & True &  &  &  &  &  &  &  \\ \cmidrule(l){2-11} 
 & QPE \cite{qpe} & \#counting qubits* & [2, \(\infty\)] & 6660 & 19 & [3, 21] & [24, 4195104] & [17, 3146336] & [11, 2097560] & [2, 20] \\
 &  & input phase initialization & 1/8 &  &  &  &  &  &  &  \\ \cmidrule(l){2-11} 
 & Shor's \cite{shor} & N* & (Section \ref{dataset_design}) & 17350 & 20 & [18, 30] & [22194, 122970] & [32677, 205864] & [14532, 89187] & [8, 14] \\
 &  & a & random &  &  &  &  &  &  &  \\ \midrule
Quantum & Grover's \cite{grover} & oracle width* & [2, \(\infty\)] & 11.3 & 13 & [2, 14] & [14, 73455] & [12, 24647] & [3, 40954] & [2, 14] \\
Search &  & oracle content & random &  &  &  &  &  &  &  \\
Algorithms &  & \#solution & \(2^{(\text{problem size}-2)}\) &  &  &  &  &  &  &  \\
 &  & \#iteration & optimal &  &  &  &  &  &  &  \\ \cmidrule(l){2-11} 
 & Quantum & \#counting qubits* & [2, \(\infty\)] & 4680 & 6 & [4, 14] & [251, 569271] & [200, 450687] & [125, 325933] & [2, 7] \\
 & Counting \cite{counting} & \#searching qubits & = problem size &  &  &  &  &  &  &  \\
 &  & \#solution & random &  &  &  &  &  &  &  \\ \cmidrule(l){2-11} 
 & Quantum & \#theta qubits & \(2^{(\text{problem size})}\) & 4040 & 2 & [11, 20] & [26919, 2018376] & [21874, 1583025] & [15124, 1045248] & [4, 8] \\
 & Walk \cite{walk} & \#node qubits & \(2^{(\text{problem size})}\) &  &  &  &  &  &  &  \\
 &  & \#coin qubits* & [2, \(\infty\)] &  &  &  &  &  &  &  \\
 &  & \#iterations & optimal &  &  &  &  &  &  &  \\
 &  & \#solutions & \(2^{(\text{problem size}-2)}\) &  &  &  &  &  &  &  \\ \midrule
Quantum & Quantum Key & circuit width* & [2, \(\infty\)] & 2.66 & 45 & [2, 46] & [5, 6] & [3, 87] & N/A & [4, 92] \\
Communication & Distribution \cite{key} & interception? & True &  &  &  &  &  &  &  \\ \cmidrule(l){2-11} 
Algorithms & Superdense & circuit width* & [2, \(\infty\)] & 2.57 & 45 & [2, 46] & [6, 95] & [3, 30] & [2, 90] & [2, 46] \\
 & Coding \cite{superdense} & encoding size & half of the qubits &  &  &  &  &  &  &  \\ \cmidrule(l){2-11} 
 & Quantum & \#qubit to teleport* & [2, \(\infty\)] & 11.9 & 14 & [6, 45] & [14, 14] & [16, 120] & [8, 60] & [2, 15] \\
 & Teleportation \cite{teleport} & state to teleport & random &  &  &  &  &  &  &  \\ \midrule
Variational & QAOA \cite{qaoa} & \#graph nodes* & [2, \(\infty\)] & 2.47 & 45 & [2, 46] & [6, 141] & [5, 138] & [2, 92] & [2, 46] \\
Quantum &  & type of graph & cyclic &  &  &  &  &  &  &  \\
Algorithms &  & gate parameters & random &  &  &  &  &  &  &  \\ \cmidrule(l){2-11} 
 & VQE \cite{vqe} & circuit width* & [2, \(\infty\)] & 50.5 & 45 & [2, 46] & [6, 94] & [8, 184] & [1, 1035] & [2, 46] \\
 &  & variational form repetition & 1 &  &  &  &  &  &  &  \\
 &  & variational form gate type & RY, RZ &  &  &  &  &  &  &  \\
 &  & type of entanglement & full entanglement &  &  &  &  &  &  &  \\
 &  & gate parameters & random &  &  &  &  &  &  &  \\ \cmidrule(l){2-11} 
 & VQC \cite{1611.09347} & circuit width* & [2, \(\infty\)] & 190 & 45 & [2, 46] & [9, 317] & [9, 1219] & [3, 2115] & [2, 46] \\
 &  & feature map repetition & 1 &  &  &  &  &  &  &  \\
 &  & variational form repetition & 1 &  &  &  &  &  &  &  \\
 &  & gate parameters & random &  &  &  &  &  &  &  \\ \bottomrule
\end{tabular}%
}
\end{table*}

\section{Q-Gen: Evaluation \& The Circuit Dataset} \label{dataset}

In this Section, we evaluate the characteristics and functionality of Q-gen by putting it to work. We explain the architecture and design philosophy of our quantum circuit dataset, which is generated by utilizing the plentiful generation parameters provided by Q-gen.

\subsection{Hierarchical Algorithm System}
The most intuitive way to demonstrate Q-gen’s capability as a circuit generator is to show that it can easily generate a large variety of quantum circuits. We realized that organizing the generated circuits into a well-structured dataset and providing open-source access can greatly assist studies that require large-scale quantum circuit testing and benchmarking, or even push new research directions on AI\(+\)Quantum.

However, the quantum algorithms provided by Q-gen have different structures and targeted applications. A good implementation of Grover’s algorithm should find the marked state in fewer iterations. In contrast, a good quantum key distribution protocol should protect the message from the attacker with minimum encryption overhead. 
The complex relationships between different quantum algorithms affect how researchers evaluate and optimize the performance of their novel projects. For example, a circuit optimization tool aimed at reducing the number of \texttt{CNOT} gates should not pick the Deutsch-Jozsa algorithm or the quantum key distribution algorithm for testing, due to their low utilization rate of \texttt{CNOT} gates. For studies aimed at improving the QFT algorithm, it is also valuable to test its performance on the QPE algorithm, because it is a direct extension of the QFT algorithm.

Therefore, to facilitate understanding and promote effective use of all the available algorithms in Q-gen, we summarize them into an organized system with 5 algorithm categories based on their theoretical origin, circuit structure, and targeted application. The complete Q-gen algorithm system is visualized in Figure \ref{alg_system}, we hope this algorithm system can offer a clear and direct introduction for new researchers to dive into quantum computing.


\subsubsection{Quantum Query Algorithms}

The quantum query algorithms include the Deutsch-Jozsa algorithm, the Bernstein-Vazirani algorithm, and Simon’s algorithm.
They try to query a ``black-box'' function (oracle) to find its hidden information.
The hidden information of the oracle can be randomly generated based on \texttt{problem size} or specified as a bitstring. Since most of the quantum computation happens inside the oracle, the oracle's complexity directly affects the generated circuit's complexity.

\subsubsection{Quantum Fourier Transform Algorithms}

The quantum Fourier transform algorithms include quantum Fourier transform (QFT), quantum phase estimation (QPE), and Shor’s
algorithm. Similar to the classical Fourier transform where functions are transferred from the time domain to the frequency domain, quantum Fourier transform (QFT) converts the quantum states from the computational basis to the Fourier basis. 
They are some of the most important algorithms in quantum computing, providing fundamental building blocks for many other quantum applications.

\subsubsection{Quantum Search Algorithms}

The quantum search algorithms include Grover’s algorithm, quantum counting, and quantum walk algorithm. The quantum search algorithms can accelerate search problems on datasets with no predefined classical data structures. The quantum search circuits still involve an oracle like the quantum query algorithms, but this oracle represents the search space, followed by additional operations according to specific algorithms. 
In general, as the search space grows, the circuit complexity of the generated circuits grows exponentially in terms of gate number/circuit depth.

\subsubsection{Quantum Communication Algorithms}

The quantum communication algorithms include quantum key distribution, superdense coding, and quantum teleportation.
The quantum communication algorithms utilize different merits of quantum channels to securely and rapidly transfer information. Q-gen generates the full quantum communication process from sender to receiver. Because the message should remain protected during the process, besides demonstrating the functionality of quantum communication, the generated circuits can also be used to benchmark the fidelity of the quantum hardware.

\subsubsection{Variational Quantum Algorithms}

The variational quantum algorithms include the quantum approximate optimization algorithm (QAOA), the variational quantum
eigensolver (VQE), and the variational quantum classifier (VQC). Variational quantum algorithms combine quantum computing with classical optimization. The quantum circuit is constructed to solve for a specific ground state of a given system, and a classical optimizer iteratively optimizes the angles of the rotation gates involved in the circuit. Note that a variational quantum circuit is often called a parametrized quantum circuit, which refers to the parameterized quantum gates involved in the circuit. Q-gen's parameterization refers to the high-level generation parameters of a quantum algorithm.

\Figure[t](topskip=0pt, botskip=0pt, midskip=0pt)[width=0.86\textwidth]{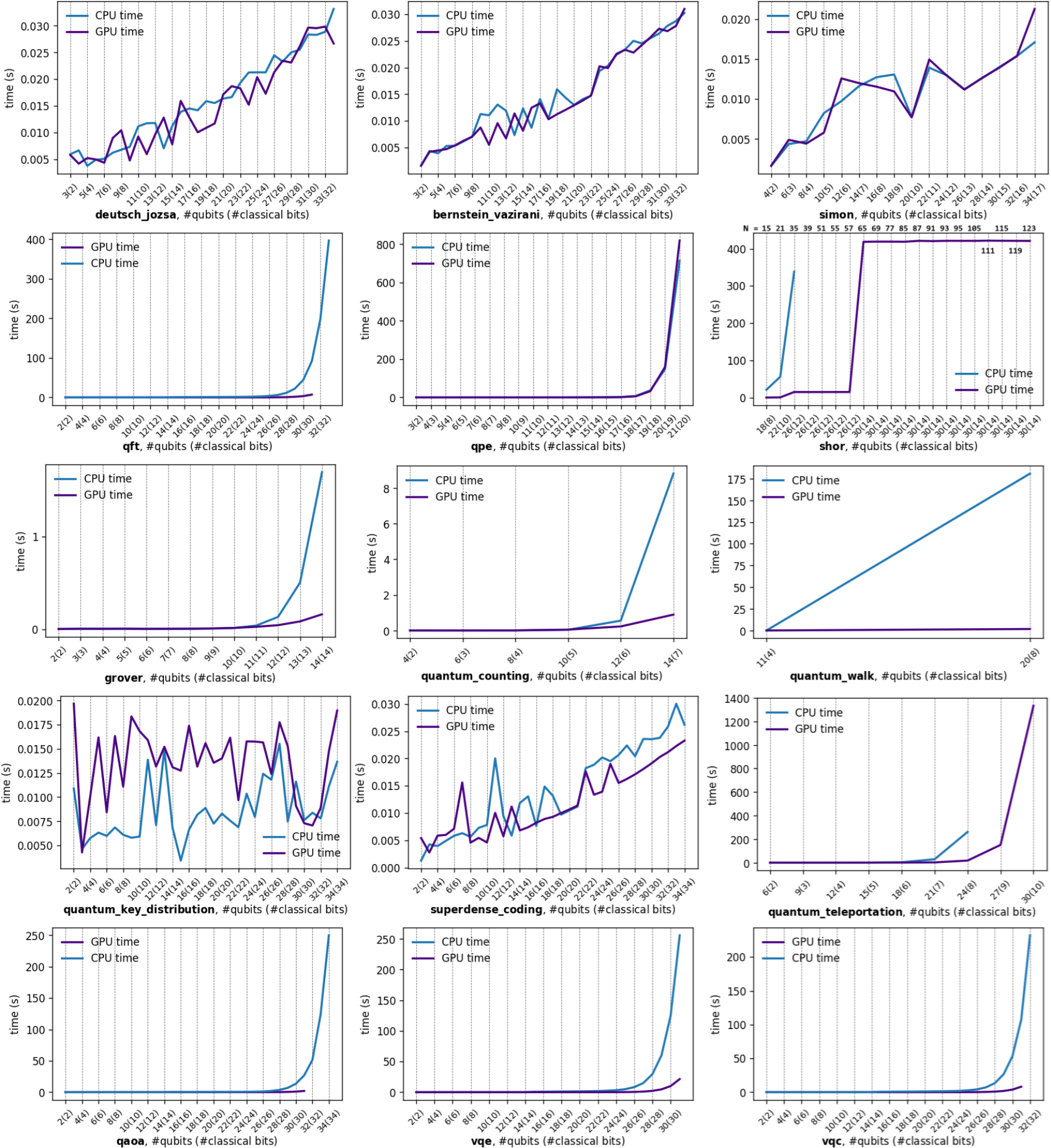} %
{Simulation times for the Q-gen database circuits using two Xeon Gold 6354 CPUs and the Nvidia A100 GPU. The x-axis shows the number of qubits and the number of classical bits (measurements) in the circuit. For Shor's algorithm, the 20 ticks represent 20 different input numbers N being solved: N = [15, 21, 35, 39, 51, 55, 57, 65, 69, 77, 85, 87, 91, 93, 95, 105, 111, 115, 119, 123]. The simulation time exhibits an exponential growth pattern for most algorithms when the number of qubits increases. For Quantum Query Algorithms, due to their simple structure indicated by the gate statistics in Table \ref{table_dataset}, the simulation time will need much more qubits to show exponential growth.\label{simulation_full}}

\subsection{Quantum Circuit Dataset Design} \label{dataset_design}

The algorithms under one Q-gen category share similar characteristics, like circuit structure and measurement output pattern. Given the high variability of the available generation parameters, it is possible to generate some outlier circuits that have unrealistic gate arrangements. Still, when creating the Q-gen dataset, we try to keep the parameters reasonable so the generated circuits are close to practical application circuits. The generation details for each algorithm category are explained below:

\begin{enumerate}
    \item The quantum query circuits are relatively easy to generate in terms of average generation time and circuit size. All the oracles in the three algorithms are set to contain random output states, and the oracle of the Deutsch-Jozsa algorithm is always set to be a constant oracle.

    \item The quantum Fourier transform circuits are significantly harder to generate. The QFT circuits are generated as inverse-QFT and measurement gates to produce verifiable histograms. The QPE circuits have fixed input initialization of \(\theta=\frac{1}{8}\), the same phase angle as the \texttt{T} gate. The input number of QFT circuits and the initial guess (\(a\)) of Shor’s circuits are both randomly generated.

    \item The quantum search circuits take the longest average time to generate in Q-gen if normalized to the same circuit width. Due to the exponentially growing gate count, it also produces the deepest circuits compared to other algorithm categories. For Grover’s algorithm and the quantum walk algorithm, we specify the number of solutions as \(2^{(\text{\texttt{problem size}}-2)}\) so that the number of iterations will always be 1. Although the number of solutions is fixed, the individual solution state inside the oracle is still randomly picked. The quantum counting circuits always generate the same number of counting qubits and searching qubits according to \texttt{problem size}.

    \item The quantum communication circuits have the simplest generation parameters in Q-gen. The quantum key distribution circuits contain interceptions from the attacker, the superdense coding circuits encode half of the qubits, and the quantum teleportation circuits randomly initialize the states to teleport.

    \item The variational quantum algorithms can produce circuits with vastly different structures depending on the generation parameters. The QAOA circuits are generated with a cyclic input graph, and the variational form of the VQE circuits is generated with \texttt{RY/RZ} gates and full entanglement. The repetition number of the variational form for the VQE and VQC circuits is fixed to 1. For all the circuits in this category, the rotation angles of the single-qubit gates are randomly generated.
\end{enumerate}

The minimum \texttt{problem size} is set to 2 for every algorithm (except Shor’s algorithm) because most of them require at least 2 qubits, it is then incremented by 1 for each generation step until the circuit gets too large for generation or simulation. For Shor’s algorithm, the circuit structure is decided by the input number to be factored. We generated Shor’s circuit with a list of \texttt{problem size} from a subset of odd composite numbers, up to 123. Table \ref{table_dataset} summarizes the full generation details and the gate statistics of the Q-gen quantum circuit dataset. 

\subsection{Algorithm Complexity Rating}

We want to provide each circuit in the Q-gen dataset with its corresponding noise-free output to form a straightforward input-output pair, either in measurement count or state vector format. This enables researchers to utilize the Q-gen dataset without worrying about the simulation overhead of large circuits, which can easily take hours to days. For example, in the use case where the user wants to train a machine learning model to perform tasks related to quantum circuits, our dataset provides not only the high-level circuits but also their corresponding simulation results. So the user does not need to run simulations by themself on these circuits, effectively reducing their overall simulation workload.

For circuit simulation, we obtain the measurement count using the Qiskit Aer simulator running on a rack server with two Xeon Gold 6354 processors and the Nvidia A100 GPU, with 512GB of shared RAM and 80GB of graphics memory. The simulation times for all the Q-gen dataset circuits are plotted in Figure \ref{simulation_full}. For most algorithms, the simulation time difference between GPU and CPU is negligible. However, the CPU simulation of Shor’s algorithm and quantum teleportation quickly ran out of memory. The GPU can significantly outperform the CPU on these two algorithms thanks to the cuQuantum SDK optimization. The state vector simulations are performed on the Qulacs simulator due to its efficiency when running large quantum circuits. The circuits are converted from Qiskit to Qulacs format with the Naniwa converter, and the simulation results can be cross-verified with the state vector output from the Aer simulator.

Another addition to complement the usability of the Q-gen quantum circuit dataset is the algorithm complexity rating, denoted by the star (\(\medwhitestar\)) symbols in Figure \ref{alg_system}. We define an algorithm’s complexity based on its simulation time, generation time, and gate statistics. Note that this rating should not be confused with the circuit complexity mentioned in Section \ref{algorithm}. A quantum circuit's complexity can be quantified regardless of its application, most commonly using the circuit depth. However, it is hard to come up with a comprehensive and uniform metric to define the complexity of a quantum algorithm across all the fields of quantum computing. Under the Q-gen algorithm category system, the individual algorithms within the same category have the same application and similar circuit structure, so we assign the ratings independently for each algorithm category. This gives researchers a clear hierarchical view of the complexity of the quantum algorithms, and helps software developers quickly grasp the coding difficulty if they want to commit new algorithms to the Q-gen algorithm system.



\section{Discussion \& Outlook} \label{discussion}

In this Section, we discuss the potential applications of the Q-gen circuit generator and the circuit dataset. We also give an outlook on the future directions of the Q-gen project.

\subsection{Practical Applications}

For machine learning applications, the size of the dataset can be expanded depending on the computing power offered by the user. On the other hand, since we provide high-level circuits in our dataset, users can utilize their own transpiler to generate more low-level circuits based on their specific learning objectives and optimization needs. For example, tuning the number of SWAP insertions or translating the circuits into different basis gate sets. Furthermore, since the original computation of the high-level circuit should be preserved during transpilation, our noise-free simulation results remain true to all low-level circuits produced. They can be used as a "truth label" to compare and analyze different transpilation techniques. For example, users can run their low-level circuits and compare their outputs with the Q-gen dataset to verify that their transpilation is accurate. Or even more creatively, if the goal is to develop an approximate transpiler that trades off some measurement accuracy for lower gate error rates, our dataset can be used as a golden reference to calculate how much the approximated output differs from the noise-free output.

Q-gen can generate circuits that solve specific problems depending on the input. For example, the user can supply any oracle to Grover’s algorithm, or define any variational form for the VQE/VQC algorithm. Q-gen will then generate the full quantum circuit as the solution to the user’s input problem. This ability to create problem-circuit pairs can be used to train an ML-based circuit generator. There are also many other practical applications with the Q-gen circuit dataset. For instance, train large models that can analyze the noise characteristics of real quantum hardware, or verify the accuracy of the results of circuit cutting/knitting where a large circuit is decomposed into smaller circuits and run separately.

\subsection{Future Improvements}

Currently, Q-gen generates high-level circuits that do not have hardware-specific generation options like setting a restricted gate set. Providing some control over how a given circuit would transpile within a given hardware environment can be a useful feature. However, most of the hardware-related transpilation parameters are unfortunately proprietary and not easily accessible to us. As we continue to refine the core generation functions of Q-gen, we will be gradually adding more generation options to the project as more open-source hardware data becomes available.

Although Q-gen does not limit the qubit number of the generated circuit, simulating extremely large circuits is still a resource-expensive task. Our work presents the initial dataset to kickstart the development of the Q-gen project, as more efficient simulating methods are being investigated, we will generate and simulate more complex circuits and add those results to the dataset as soon as they are available.

In the current stage of quantum computing where the main objective for hardware is to reduce error and achieve scalability, we believe it is more practical to set the targeted application of Q-gen as generating high-level circuits that can support the classical design and optimization methods for quantum algorithms. Nevertheless, quantum hardware is experiencing exploding development. To add more possibilities to the Q-gen dataset, we also plan to add more outputs from real quantum hardware.





\section{Conclusion} \label{conclusion}


In this work, we present the Q-gen quantum circuit generator to intuitively and efficiently generate practical quantum circuits. We motivate this work by showing that a comprehensive quantum circuit generator can accelerate the advancement of current NISQ applications, and may become an integral part of the future quantum computing workflow. Q-gen provides 15 practical quantum algorithms with tailored generation parameters, offering a convenient tool for researchers and developers with any background level to quickly start interacting with quantum circuits. We demonstrate the functionality of Q-gen by generating a large quantum circuit dataset utilizing all the available generation parameters, and we evaluate the structure of the generated circuits, showing a wide coverage of different circuit properties including circuit size and gate counts. Based on the circuit dataset and its simulation results, we further improve the usability of Q-gen by defining a hierarchical quantum algorithm system, presenting a streamlined starting point for quantum computing research.


\section{Data \& Code Availability}

The source code of the Q-gen quantum circuit generator is hosted on GitHub \cite{github}. We provide the generation functions for every algorithm, as well as the dataset generation program to directly generate a full quantum circuit dataset. An example of generating a Deutsch-Jozsa circuit is given below:

\begin{lstlisting}[language=Python]
n = 4
oracle = 'balanced'
print_oracle = True

# n = number of qubits of the oracle
# options = [oracle, print_oracle]
# oracle = 'constant' (easy), 'balanced' (hard)
# print_oracle = T/F
dj_qc = deutsch_jozsa(n, [oracle, print_oracle])
\end{lstlisting}

We host all the supplementary data related to the Q-gen dataset and the generator on the dedicated GitHub Wiki page \cite{github_wiki}. We provide high-quality circuit visualizations, as well as simulation result plots showing the measurement counts. We will also update other helpful data related to the dataset as soon as they become available. For example, the output and the transpilations of our Q-gen dataset circuits running on real IBM Quantum processors \cite{ibmq}.

The Q-gen quantum circuit dataset is hosted on Kaggle\cite{kaggle}. We provide the Qiskit/Qulacs circuits and the simulation result files. Examples of loading the circuit and simulation result are given below:

\begin{lstlisting}[language=Python]
algorithm = ['deutsch_jozsa', 'bernstein_vazirani',
    'simon', 'grover', 'quantum_counting',
    'quantum_walk', 'qft', 'qpe', 'shor', 
    'quantum_key_distribution',
    'superdense_coding', 'quantum_teleportation', 
    'qaoa', 'vqe', 'vqc']

# algorithm index
i = 0
# circuit index
j = 0

# load .qpy circuits
with open('circuits_qiskit/' + f"{i:02}_"
          + algorithm[i] + '_circuits.qpy', 
          'rb') as f:
    qpy_circuit_list = qpy.load(f)
    qpy_circuit = qpy_circuit_list[j]

# load qulacs circuits
with open('circuits_qulacs/' + f"{i:02}_"
          + algorithm[i] + '_qulacs.pickle',
          'rb') as f:
    qulacs_circuit_list = pickle.load(f)
    qulacs_circuit = qulacs_circuit_list[j]

# load Qiskit simulation result object
with open('sim_measurement_xxx/' + f"{i:02}_"
          + algorithm[i] + '_sim_result.pickle', 
          'rb') as f:
    results = pickle.load(f)

# load Qiskit state vector simulation output
with gzip.open('sim_statevector_qiskit/'
               + f"{i:02}_" + algorithm[i]
               + '/sim_result_' + f"{j:02}"
               +'.txt.gz', 'r') as f:
    for line in f:
        print(line.decode()[:-1] + '  ')

# load qulacs state vector simulation output
sim_sv = np.loadtxt('sim_statevector_qulacs/'
                    + f"{i:02}_" + algorithm[i]
                    + '/sim_result_'
                    + f"{j:02}.txt.gz",
                    dtype = np.complex_)
\end{lstlisting}


\section*{Acknowledgment}


We thank the Qiskit community for sharing the algorithm implementations on the now deprecated Qiskit Textbook\cite{qiskitextbook}. Q-gen is an open-source project, the generated circuits are experimental and should be used for research purposes only. 


\bibliographystyle{IEEEtran}
\bibliography{IEEEabrv.bib,IEEEexample.bib,refs.bib}{}

\newpage
\begin{IEEEbiography}[{\includegraphics[width=1in,height=1.25in,clip,keepaspectratio]{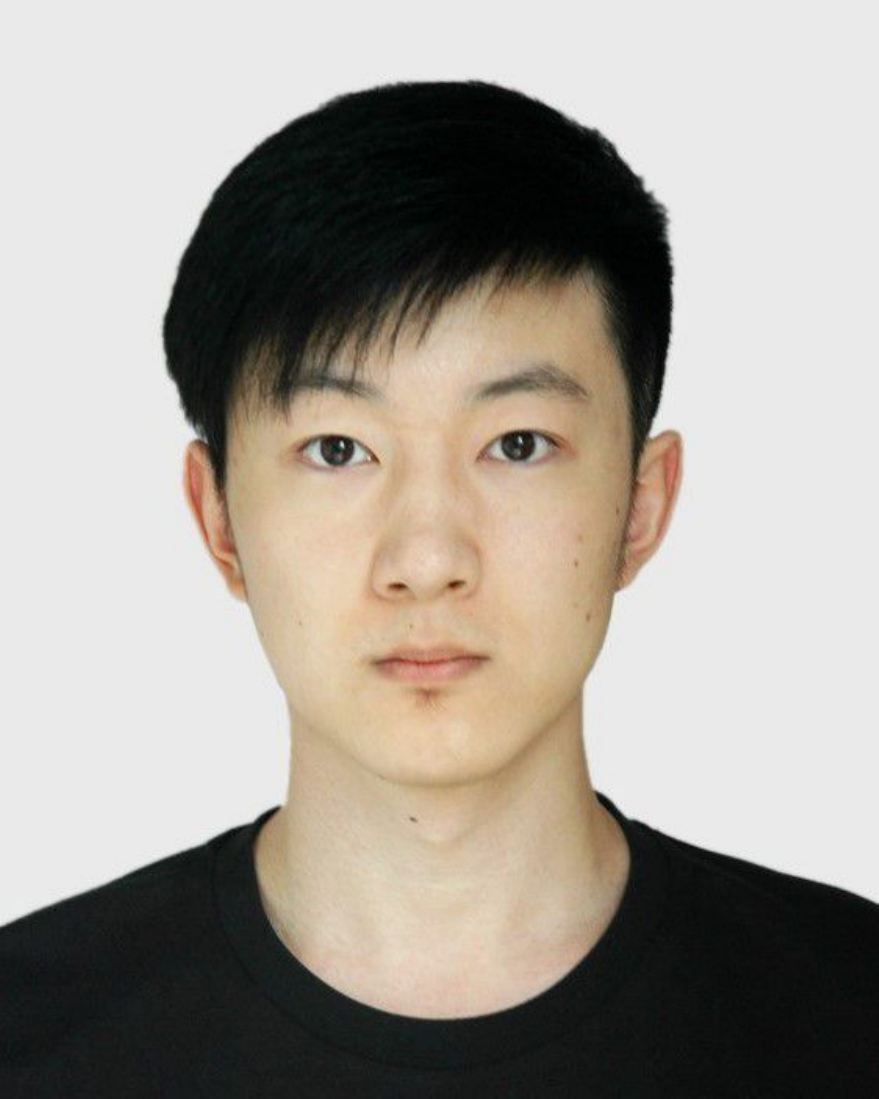}}]{Yikai Mao} is a Ph.D. student at Kondo Lab, from the Graduate School of Science and Technology, Keio University. He received his bachelor’s degree in computer engineering from the University of Florida in 2018, and his master’s degree in electrical and computer engineering from the University of California, Davis in 2021. His current research interests include quantum computing, quantum-classical hybrid computing, and computer architecture.
\end{IEEEbiography}
\vskip 0pt plus -1fil
\begin{IEEEbiography}[{\includegraphics[width=1in,height=1.25in,clip,keepaspectratio]{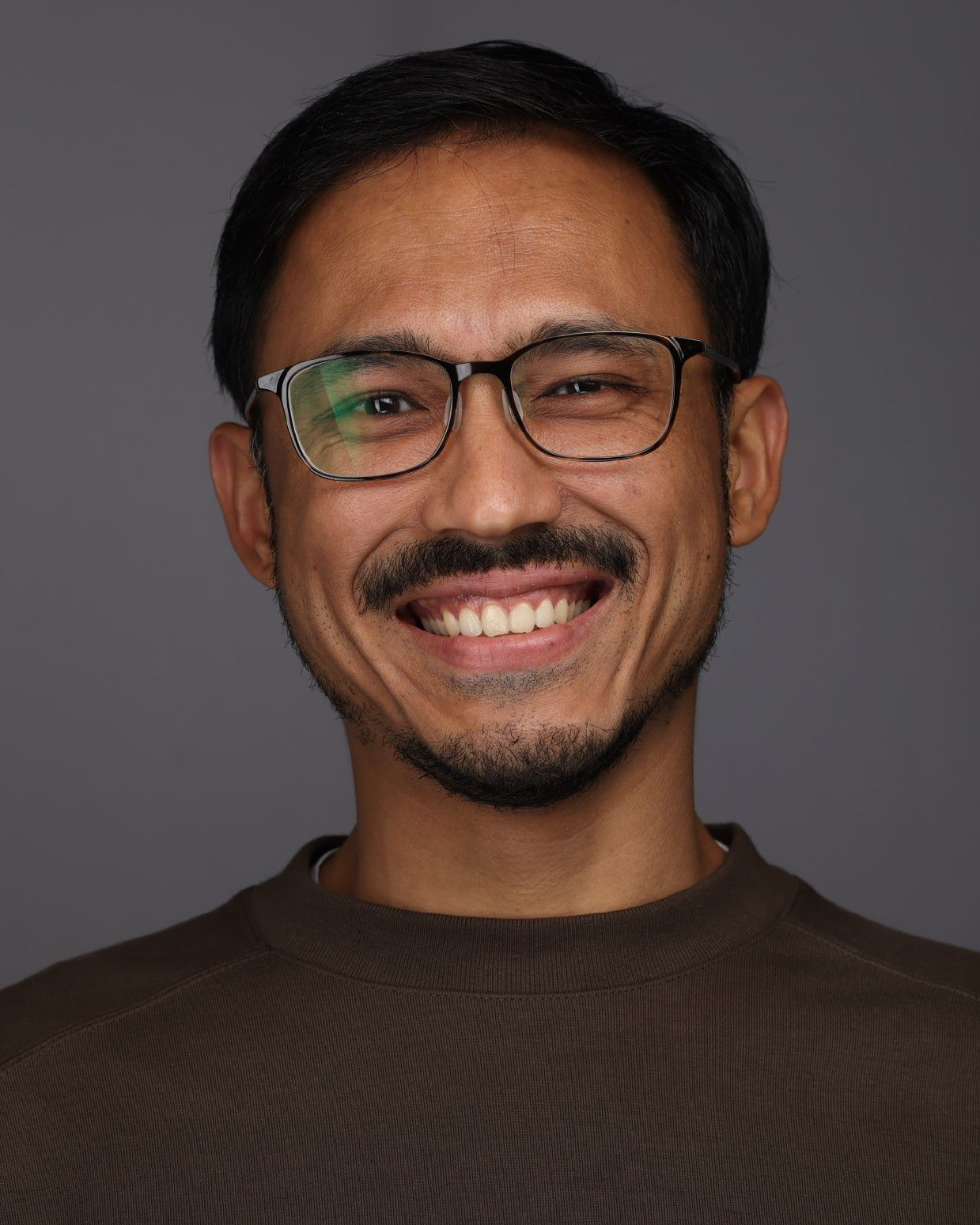}}]{Shaswot Shresthamali} is a Research Associate Professor in the Department of Advanced Information Technology at Kyushu University. He earned his doctorate from The University of Tokyo in 2021 where his research focused on resource management and scheduling in IoT nodes with Reinforcement Learning. He is currently affiliated with the Cyber-Physical Computing laboratory at Kyushu University. His research interests span the intersections of Machine Learning, Novel Computing Architectures, and Quantum Computing. He is also a Visiting Researcher at Keio University.
\end{IEEEbiography}
\vskip 0pt plus -1fil
\begin{IEEEbiography}[{\includegraphics[width=1in,height=1.25in,clip,keepaspectratio]{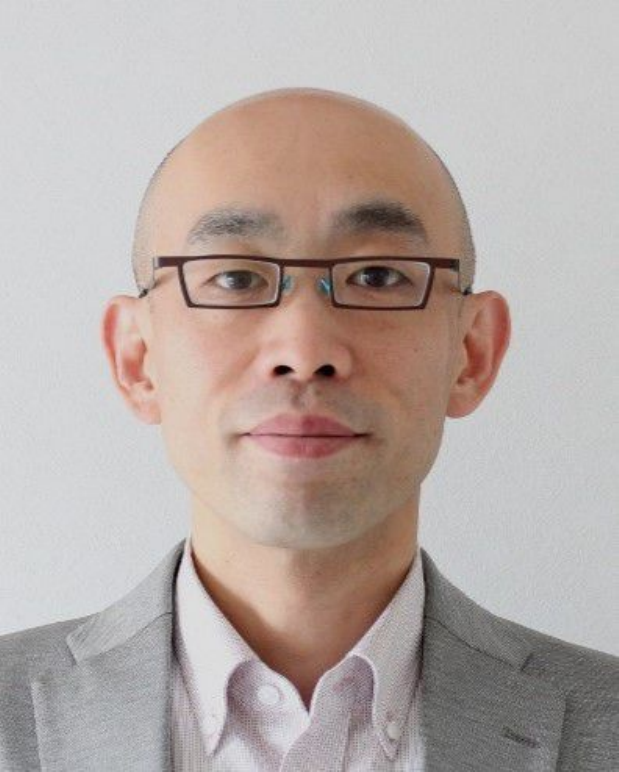}}]{Masaaki Kondo} received Ph.D in Engineering from the University of Tokyo in 2003. He is currently a professor of Faculty of Science and Technology, Keio University, and also the team leader of next generation high performance architecture research team in the RIKEN Center for Computational Science. His research interests include computer architecture, high-performance computing, quantum computer, and low-power LSI design.
\end{IEEEbiography}

\EOD

\end{document}